\documentclass{article}
\usepackage{eurobert_conference}

\usepackage{microtype}
\usepackage{hyperref}

\usepackage{url}

\usepackage{booktabs}
\usepackage{graphicx}
\usepackage{subfigure}
\usepackage{array}
\usepackage{amsmath}
\usepackage{lineno}
\usepackage{pifont}
\usepackage{multirow}
\usepackage{arydshln}
\usepackage{tikz}
\usepackage{enumitem}
\usepackage{pgfplots}
\usepackage{float}
\usepackage{caption}
\usepackage{subcaption}
\usepackage{tablefootnote}

\usepackage{tabularx}

\usepgfplotslibrary{groupplots}
\usepgfplotslibrary{fillbetween}
\pgfplotsset{compat=1.18}
\usepackage{adjustbox}

\usetikzlibrary{pgfplots.groupplots}
\usetikzlibrary{tikzmark, fit}
\usepackage[normalem]{ulem}

\definecolor{darkblue}{rgb}{0, 0, 0.5}
\hypersetup{colorlinks=true, citecolor=darkblue, linkcolor=darkblue, urlcolor=darkblue}

\definecolor{MainColour}{RGB}{130,11,46}
\definecolor{ComplementaryColour}{RGB}{11,130,95}

\usepackage[most]{tcolorbox}

\newtcbox{\clustertab}[1]{on line, box align=base, colback={#1},colframe={#1},size=fbox,arc=2pt,top=-1.5pt, bottom=-1.5pt, left=-1.5pt, right=-1.5pt, boxrule=0pt, enlarge left by=1pt}

\newtcbox{\smallclustertab}[1]{on line, box align=base, colback={#1},colframe={#1},size=fbox,arc=2pt,top=-1.2pt, bottom=-1.5pt, left=-1.0pt, right=-1.0pt, boxrule=0pt, enlarge left by=1.2pt}

\usepackage{arydshln}
\usepackage{booktabs}                     
\usepackage{array}                        

\makeatletter
\def\adl@drawiv#1#2#3{%
        \hskip.5\tabcolsep
        \xleaders#3{#2.5\@tempdimb #1{1}#2.5\@tempdimb}%
                #2\z@ plus1fil minus1fil\relax
        \hskip.5\tabcolsep}
\newcommand{\cdashlinelr}[1]{%
  \noalign{\vskip 1.3pt
           \global\let\@dashdrawstore\adl@draw
           \global\let\adl@draw\adl@drawiv}
  \cdashline{#1}[.4pt/2pt]
  \noalign{\global\let\adl@draw\@dashdrawstore
           \vskip 3pt}}
\makeatother

\title{\textsc{ViDoRe Benchmark V2}:\\ Raising the Bar for Visual Retrieval}

\author{Quentin Macé$^{1}$ \quad António Loison$^{1}$ \quad Manuel Faysse $^{1,2}$\\\small{$^{1}$Illuin Technology $^2$CentraleSupélec \,\\ } 
}

\abstract{The ViDoRe Benchmark V1 was approaching saturation with top models exceeding 90\% nDCG@5, limiting its ability to discern improvements. ViDoRe Benchmark V2 introduces realistic, challenging retrieval scenarios via blind contextual querying, long and cross-document queries, and a hybrid synthetic and human-in-the-loop query generation process. It comprises four diverse, multilingual datasets and provides clear evaluation instructions. Initial results demonstrate substantial room for advancement and highlight insights on model generalization and multilingual capability. This benchmark is designed as a living resource, inviting community contributions to maintain relevance through future evaluations.}

\contact{quentin.mace@illuin.tech}
\website{https://huggingface.co/blog/manu/vidore-v2}
\leaderboard{https://huggingface.co/spaces/vidore/vidore-leaderboard}
\date{March 18, 2025}

\definecolor{lightgray}{gray}{0.9} 

\begin{document}
\maketitle

\section*{Why a new benchmark?}
Since the release of the original ViDoRe Benchmark \citep{faysse2025colpaliefficientdocumentretrieval}, evaluating visual models on document retrieval tasks, visual retrieval models have significantly advanced! While the original ColPali model reported an average score of 81.3 nDCG@5, current SOTA models on the leaderboard surpass a nDCG@5 of 90, with some tasks becoming “too easy” to yield a meaningful signal!
With the benchmark approaching saturation for SOTA models, there is limited room to truly measure improvements and understand model capabilities in realistic scenarios. To continue pushing the boundaries of visual retrieval, it became essential to introduce a new benchmark designed specifically to challenge these advanced models: \textsc{ViDoRe Benchmark V2}.

\section{Motivating the Creation of ViDoRe Benchmark V2}
In developing ViDoRe Benchmark V2, our main goal was to create a benchmark reflective of real-world retrieval challenges—difficult, diverse, and meaningful. Current benchmarks exhibit limitations that prevent them from accurately reflecting real user behavior and complex retrieval scenarios \citep{thakur2025freshstackbuildingrealisticbenchmarks}. We identified three critical issues in existing benchmarks:

\begin{enumerate}
  \item \textbf{Extractive Nature of Queries:} Current benchmarks typically rely on extractive queries, providing unrealistic retrieval contexts since real users rarely formulate queries from exact phrases in documents.
  \item \textbf{Single-Page Query Bias:} Many benchmarks overly emphasize retrieval from single-page contexts, neglecting complex, multi-document or cross-document queries common in real-world applications.
  \item \textbf{Challenges in Synthetic Query Generation:} Purely synthetic benchmarks, while appealing in theory, are difficult to implement effectively without extensive manual oversight. They often produce outliers, irrelevant or trivial queries, making human filtering essential yet costly.
\end{enumerate}

\section{Design Decisions and Techniques Used}
To address these challenges and create a robust, realistic benchmark, ViDoRe Benchmark V2 includes several innovative features:

\begin{itemize}
  \item \textbf{Blind Contextual Querying:} In practice, users don’t often know the content of the corpus they are querying. To reduce the widespread extractive bias in most synthetic query-document datasets (datasets are often created with knowledge of the document content), we only provided query annotator models with limited information about the document (summaries, metadata, etc) and filtered out the many irrelevant queries that resulted, better reproducing real-world user interactions with the corpus.
  \item \textbf{Long and Cross-Document Queries:} Unlike traditional benchmarks, ViDoRe Benchmark V2 emphasizes long-form and cross-document queries, closely mirroring real-world retrieval situations. Multiple datasets specifically focus on scenarios involving comprehensive documents or multi-document retrieval tasks.
  \item \textbf{Hybrid Synthetic and Human-in-the-Loop Creation:} Recognizing the limitations of synthetic query generation alone, we adopted a hybrid approach—generating queries synthetically and extensively refining them through human review. This process, though intensive, ensured significantly higher query quality and dataset reliability.
\end{itemize}

\section{Dataset Selection for ViDoRe Benchmark V2}
The selected datasets (\autoref{tab:compact-dataset-summary}) for ViDoRe Benchmark V2 are diverse, publicly available, and challenging. Each dataset presents distinct visual complexity and is suitable for realistic retrieval tasks, including multilingual versions with queries translated into French, English, Spanish, and German. This multilingual approach further extends the applicability and challenge level of the benchmark.
Each dataset is associated to a multilingual version with translated queries.

\begin{table}[ht]
\centering
\scriptsize
\setlength{\tabcolsep}{5pt}      
\renewcommand{\arraystretch}{1.1} 
\begin{tabular}{@{} 
  l               
  c c c c c c c c   
  p{5cm}        
@{}}

\textbf{Dataset} & 
\rotatebox{90}{\textbf{Orig.\ Lang}} & 
\rotatebox{90}{\textbf{Query Lang}} & 
\rotatebox{90}{\textbf{\# Unique Docs}} & 
\rotatebox{90}{\textbf{\# Pages}} & 
\rotatebox{90}{\textbf{Query Subset}} & 
\rotatebox{90}{\textbf{\# Queries}} & 
\rotatebox{90}{\textbf{\# Qrels}} & 
\rotatebox{90}{\textbf{Avg.\ Pages/Query}} & 
\textbf{Comments} \\
\midrule
Insurance Terms of Service\tablefootnote{Since the dataset release, the insurance dataset was removed from the dataset for legal copyright reasons.} & Fr & Fr  &  4 &  260 & - &  18 &  86 & 4.8 & Small but challenging, multi-document \\
Biomedical & En & En & 27 & 1,016 & - & 160 & 515 & 3.2 & Largest dataset, most extractive \\
Economics & En & En &  5 & 452 & - & 58 &  907 & 15.6 & Cross-document queries, high complexity \\
\multirow{2}{*}{ESG Reports}          & \multirow{2}{*}{En} & \multirow{2}{*}{En}  & \multirow{2}{*}{30} & \multirow{2}{*}{1,538}   & Synthetic &  57 & 222 & 3.9 & \multirow{2}{*}{Natively cross-lingual, industry-specific} \\ 
            &  &   &  &  & Human & 52 & 128 & 2.5 & \\
\bottomrule
\end{tabular}
\caption{Summary of dataset statistics. Feel free to explore datasets on HuggingFace.
}
\label{tab:compact-dataset-summary}
\end{table}


\section{Evaluating Models}
To evaluate models on ViDoRe Benchmark 2, we follow these steps:

\subsection*{Option 1: Using the CLI}
Here is a CLI example for using a colpali type retriever on vidore benchmark 2. For other retrievers, please refer to this repo.

\begin{verbatim}
vidore-benchmark evaluate-retriever \\
    --model-class colpali \\
    --model-name vidore/colpali-v1.3 \\
    --collection-name vidore/vidore-benchmark-v2-dev-67ae03e3924e85b36e7f53b0 \\
    --dataset-format beir \\
    --split test
\end{verbatim}

\subsection*{Option 2: Creating a custom retriever}
Detailed instructions on how to create a custom retriever are available at \url{https://github.com/illuin-tech/vidore-benchmark}. We will soon transition to using the MTEB \citep{muennighoff_mteb_2022} library to evaluate all models.

\section{Results}
Here are for example some nDCG@5 results of visual retrieval models on ViDoRe Benchmark 2 \citep{faysse2025colpaliefficientdocumentretrieval,  ma2024unifyingmultimodalretrievaldocument, zhang2024gme, yu2024visragvisionbasedretrievalaugmentedgeneration, nomicembedmultimodal2025}.\footnote{We adapted the evaluation procedure for the voyageAI API, resulting in slightly lower performance on the ViDoRe benchmark v1 compared to the values reported by voyageAI. This discrepancy likely arises from our resizing of input images to a maximum image height of 1200 pixels to facilitate efficient benchmarking, a preprocessing step presumably not applied in voyageAI's original benchmarking setup.}

\begin{table}[ht]
\centering
\scriptsize
\setlength{\tabcolsep}{2pt}
\renewcommand{\arraystretch}{1}
\begin{tabular}{@{}l*{9}{c}|l}
\toprule
\textbf{Model} &
\textbf{\rotatebox{90}{ESG Reports (Manual)}} & 
\textbf{\rotatebox{90}{Insurance}} & 
\textbf{\rotatebox{90}{Insurance Multilingual}} & 
\textbf{\rotatebox{90}{Economics}} & 
\textbf{\rotatebox{90}{Biomedical}} & 
\textbf{\rotatebox{90}{Bio Multilingual}} & 
\textbf{\rotatebox{90}{ESG Reports}} & 
\textbf{\rotatebox{90}{ESG Reports Multilingual}} & 
\textbf{\rotatebox{90}{Economics Multilingual}} & 
\textbf{\rotatebox{90}{Average}} \\
\midrule
voyageai & 0.561 & 0.641 & 0.595 & 0.588 & 0.564 & 0.515 & 0.472 & 0.462 & 0.550 & 0.550 \\
metrics-AI/colqwen2.5-3B & 0.645 & 0.579 & 0.557 & 0.566 & 0.639 & 0.569 & 0.496 & 0.492 & 0.535 & 0.564 \\
colsmolvlm-v0.1 & 0.624 & 0.555 & 0.432 & 0.609 & 0.581 & 0.505 & 0.511 & 0.476 & 0.474 & 0.530 \\
colqwen2-v1.0 & 0.622 & 0.651 & 0.572 & 0.615 & 0.618 & 0.565 & 0.534 & 0.542 & 0.532 & 0.583 \\
colpali-v1.2 & 0.321 & 0.560 & 0.458 & 0.531 & 0.585 & 0.557 & 0.519 & 0.540 & 0.479 & 0.505 \\
dse-qwen2-2b-mrl-v1 & 0.614 & 0.655 & 0.563 & 0.615 & 0.592 & 0.551 & 0.549 & 0.557 & 0.528 & 0.580 \\
colSmol-256M & 0.460 & 0.504 & 0.341 & 0.534 & 0.532 & 0.340 & 0.272 & 0.313 & 0.273 & 0.397 \\
colpali-v1.3 & 0.511 & 0.598 & 0.501 & 0.516 & 0.597 & 0.565 & 0.570 & 0.557 & 0.499 & 0.546 \\
colqwen2.5-v0.2 & 0.684 & 0.603 & 0.532 & 0.598 & 0.636 & 0.611 & \textbf{0.574} & \textbf{0.574} & \textbf{0.565} & 0.597 \\
dse-llamaindex & 0.631 & 0.688 & \textbf{0.610} & 0.612 & 0.606 & 0.569 & 0.503 & 0.512 & 0.528 & 0.584 \\
tsystems/colqwen2.5-3b-multi-v1.0 & \textbf{0.721} & \textbf{0.693} & 0.600 & 0.548 & \textbf{0.653} & \textbf{0.617} & 0.517 & 0.533 & 0.512 & \textbf{0.599} \\
gme-qwen2-VL-7B & 0.658 & 0.607 & 0.554 & \textbf{0.629} & 0.640 & 0.551 & 0.543 & 0.567 & 0.562 & 0.590 \\
visrag-ret & 0.537 & 0.505 & 0.452 & 0.596 & 0.548 & 0.477 & 0.459 & 0.464 & 0.487 & 0.503 \\
colSmol-500M & 0.522 & 0.587 & 0.377 & 0.503 & 0.543 & 0.421 & 0.392 & 0.391 & 0.361 & 0.455 \\
colpali-v1.1 & 0.465 & 0.547 & 0.484 & 0.567 & 0.564 & 0.507 & 0.461 & 0.481 & 0.438 & 0.502 \\
\bottomrule
\end{tabular}
\caption{Model performance across datasets (nDCG@5). Highest per column in \textbf{bold}.}
\label{tab:model-results-transposed}
\end{table}


\begin{figure}[ht]
    \centering
    \includegraphics[width=\textwidth]{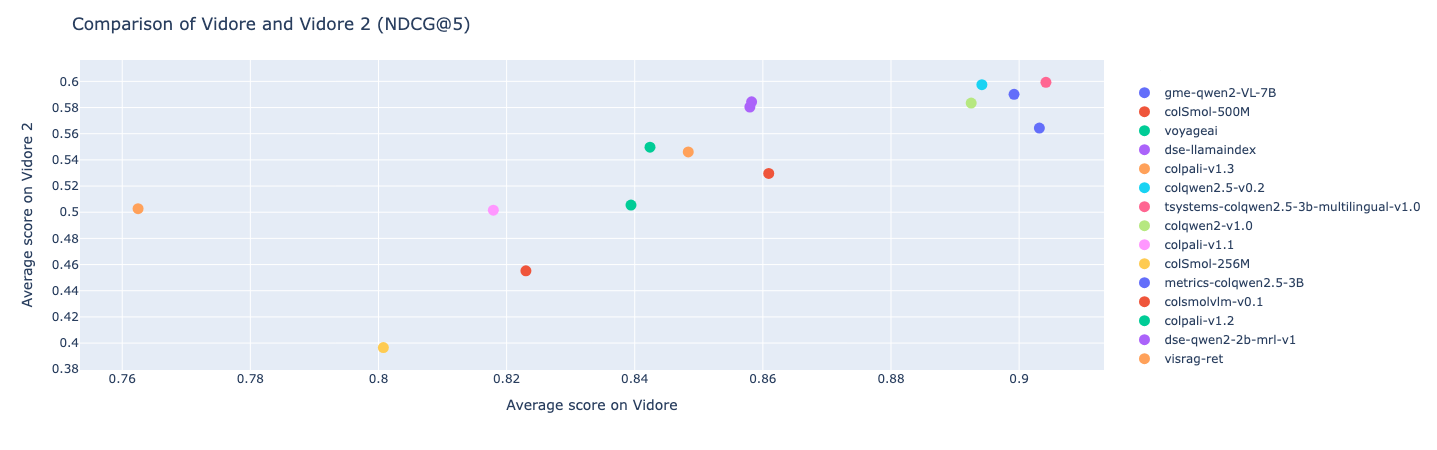}
    \caption{Performance results across models for V1 and V2. We observe strong correlations, although a clear saturation on V1 for top models. Results are in nDCG@5.}
    \label{fig:v1vsv2}
\end{figure}

\begin{figure}[ht]
    \centering
    \includegraphics[width=\textwidth]{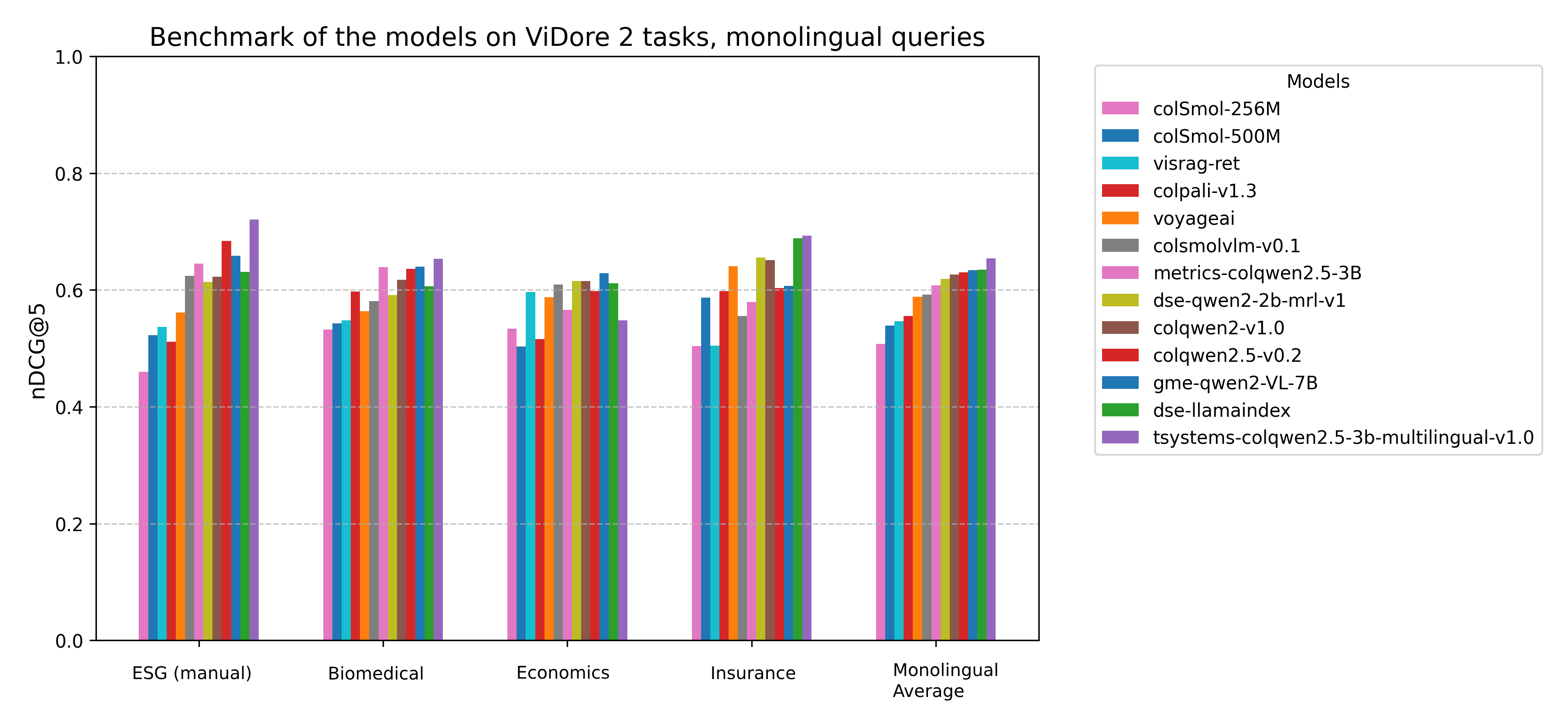}
    \caption{Performance results across monolingual tasks. ViDoRe v2 leaves substantial room for future improvements, contrasting with ViDoRe v1, which was approaching performance saturation. }
    \label{fig:monolingual}
\end{figure}

\begin{figure}[ht]
    \centering
    \includegraphics[width=\textwidth]{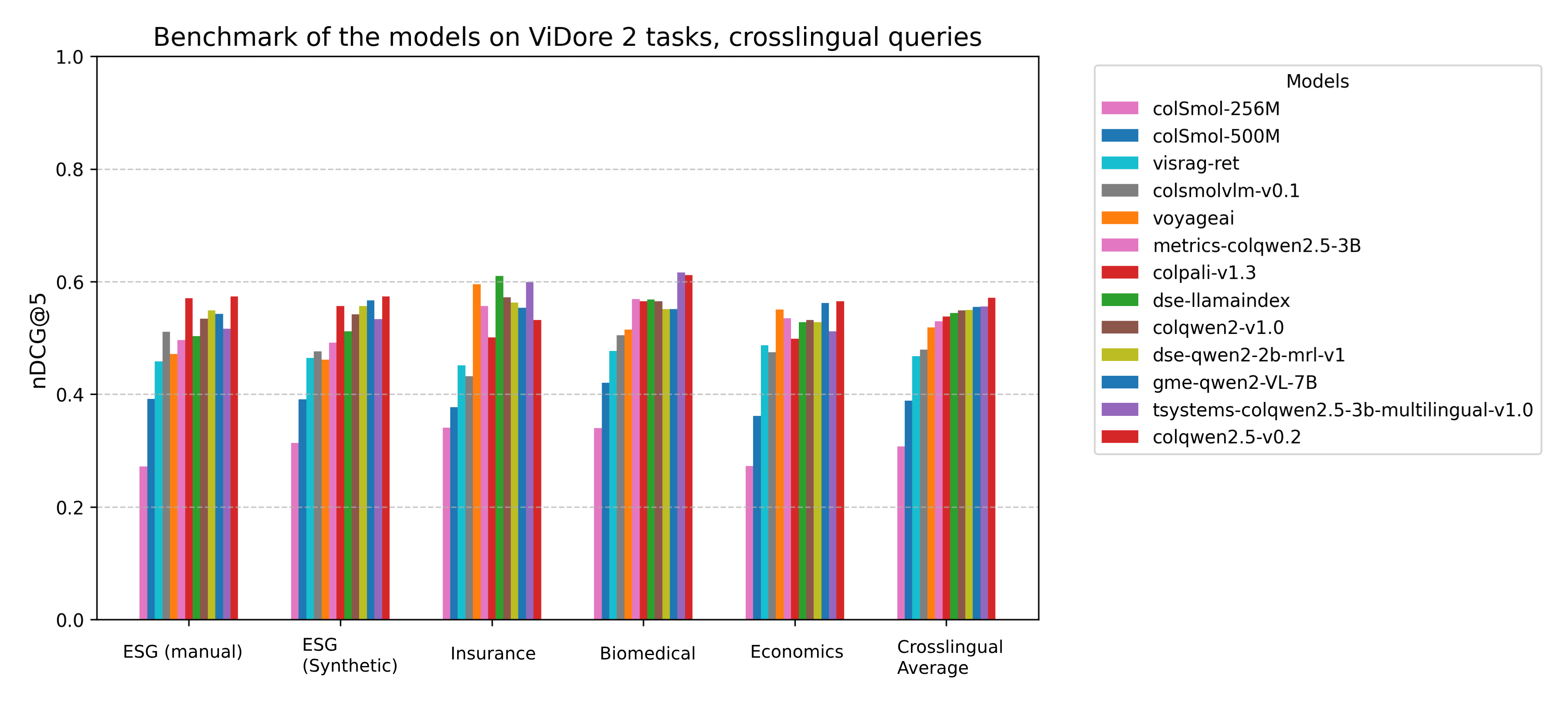}
    \caption{Performance results across crosslingual tasks.We observe a significant performance gap between models trained exclusively in English using an English-only VLM and those that are not.}
    \label{fig:multilingual}
\end{figure}

Analyzing the results, we can extract a few general takeaways:

\begin{itemize}
  \item ViDoRe v2 maintains strong correlation with v1, with consistent model rankings across versions (\autoref{fig:v1vsv2}). 
  \item ViDoRe v2 leaves substantial room for future improvements, contrasting with ViDoRe v1, which was approaching performance saturation (scores exceeding 90\% as seen in \autoref{fig:monolingual}).
  \item Certain models exhibit signs of slightly overfitting to the training distribution, resulting in reduced generalization to novel data (e.g., vidore/colSmol-256M, vidore/colSmol-500M, Metric-AI/ColQwen2.5-3b-multilingual-v1.0). These models perform worst on the V2 than what their performance on the V1 would lead to believe (\autoref{fig:v1vsv2}).
  \item The multilingual splits in ViDoRe v2 provide a more accurate assessment of non-english capabilities in visual retriever models. We observe a significant performance gap between models trained exclusively in English using an English-only VLM and those that are not.(\autoref{fig:multilingual})
  \item Larger model scale is beneficial; notably, the gme-qwen7B model achieves strong overall performance but incurs significant computational cost and inference latency. Inversely, while impressive for their sizes, models under 1B parameters tend to lag behind, especially on previously unseen data distributions.
  \item We tend to see better separation between model performances with the human labeled dataset (ESG human), indicating it is of slightly higher quality than the synthetic datasets and is a more discriminating signal (\autoref{fig:monolingual}).
\end{itemize}

\section{Moving Forward}
Our goal is for ViDoRe V2 to become a dynamic, "living benchmark" that regularly grows with new tasks and datasets. To achieve this, we welcome and encourage the community to contribute datasets and evaluation tasks. This collaborative approach helps ensure that the benchmark stays relevant, useful, and reflective of real-world challenges.

We are also open on integrating new retrieval metrics such as confidence estimation measures \citep{gisserotboukhlef2024trustworthyrerankingsimpleeffective}, increasing multilingual coverage allowed by ever better base models \citep{yang2025qwen251mtechnicalreport, boizard2025eurobertscalingmultilingualencoders}, and extending the leaderboard to new modalities (audio, image querying, etc...)

\section*{Acknowledgements}

Training compute for running evaluations is obtained on the Jean Zay supercomputer operated by GENCI IDRIS through compute grant AD011016393.

For deeper discussions and projects around Visual RAG, ColPali, or agentic systems, please contact \href{mailto:contact@illuin.tech}{contact@illuin.tech} or visit \url{https://www.illuin.tech}. We welcome community contributions of document-query sets to enhance this living benchmark.

\newpage

\bibliography{eurobert_conference}
\bibliographystyle{eurobert_conference}

\newpage

\end{document}